\documentstyle[twoside,fleqn,espcrc2]{article}

\newcommand{\AmS}{{\protect\the\textfont2
  A\kern-.1667em\lower.5ex\hbox{M}\kern-.125emS}}
\begin{document}

\title{Improving QCD with fermions: the 2 dimensional case}

\author{Xiang-Qian Luo
\address{Department of Physics,
       Zhongshan University, Guangzhou 510275, 
       China},
Jun-Qin Jiang
\address{Department of Physics, Guangdong Institute of Education, 
Guangzhou 510303, China},
Shuo-Hong Guo $^{\rm{a}}$, Jie-Ming Li $^{\rm{a}}$, 
Jin-Ming Liu $^{\rm{a}}$, Zhong-Hao Mei$^{\rm{a}}$,
Hamza Jirari $^{\rm{c}}$, Helmut Kr\"oger
\address{D\'epartement de Physique, Universit\'e Laval, 
Qu\'ebec, Qu\'ebec G1K 7P4, Canada},
Chi-Min Wu
\address{Institute for High Energy Physics, 
Academia Sinica, Beijing 100039, China}
}

\begin{abstract}
We study QCD in 2 dimensions using
the improved lattice fermionic Hamiltonian 
proposed by Luo, Chen, Xu and Jiang. 
The vector mass and the chiral condensate 
are computed for various $SU(N_C)$ gauge groups. 
We do observe considerable improvement 
in comparison with the Wilson quark case.
\end{abstract}

\maketitle


\section{INTRODUCTION}

Wilson's lattice quark formulation has $O(a)$ errors, 
inducing systematic uncertainties 
when extracting continuum physics. 
The most efficient way for reducing 
these errors is the Symanzik improvement. 
There have been several reasonable proposals:

\noindent
(a) Hamber and Wu proposed the first improved action \cite{Hamber83}
to remove the $O(a)$ error. 

\noindent
(b) 
Correspondingly,
Luo, Chen, Xu, and Jiang constructed an improved Hamiltonian 
\cite{Luo94}. 
	
\noindent
(c) For other proposals, see Refs. \cite{Sheik85}.

There have been troubles in the 
Lagrangian (action) formulation: i.e., it is extremely
difficult to study $S$-matrix and cross sections, 
wave functions of vacuum, hadrons and glueballs,
QCD at finite baryon density, 
or the computation of QCD structure functions 
in the region of small $x_B$ and $Q^2$. 

The Hamiltonian approach is a viable alternative \cite{Kroger92}
and some very interesting results 
\cite{Luo96_2,Kroger98,Schutte97}
have recently been obtained.
Workers in Lagrangian formulation nowadays have followed similar ideas 
by considering anisotropic lattices to improve the spectrum computation. 

The purpose of this work is to show that in the case of 
$\rm{QCD}_2$, the improved Hamiltonian theory proposed
by Luo, Chen, Xu, and Jiang \cite{Luo94} 
can significantly reduce 
the $O(a)$ errors.

\section{IMPROVED QCD WITH QUARKS}

The fermionic Hamiltonian \cite{Luo91_1} with Wilson quarks with $O(a)$ errors is

\begin{eqnarray*}
H_{f}=  m\sum_{x} \bar {\psi}(x) \psi (x) 
\end{eqnarray*}
\begin{eqnarray*}
+
{1 \over 2a} \sum_{x,k} \bar{\psi}(x) \gamma_{k} U(x,k) \psi (x+k)
\end{eqnarray*}
\begin{eqnarray}
+ {r \over 2a} \sum_{x,k}[\bar{\psi}(x) \psi (x)-
\bar{\psi} (x) U(x,k) \psi (x+k)],
\label{unimprovedH}
\end{eqnarray}

Luo, Chen, Xu and Jiang's improved Hamiltonian \cite{Luo94} is

\begin{eqnarray*}
H_{f}^{improved}= m\sum_{x} \bar {\psi}(x) \psi (x) 
+{r \over 2a} \sum_{x,k}\bar{\psi}(x) \psi (x)
\end{eqnarray*}
\begin{eqnarray*}
+
{b_{1} \over 2a} \sum_{x,k}\bar{\psi} (x) \gamma_{k} U(x,k) \psi (x+k)
\end{eqnarray*}
\begin{eqnarray*}
+{b_{2} \over 2a}\sum_{x,k}\bar{\psi} (x) \gamma_{k} U(x,2k) \psi (x+2k)
\end{eqnarray*}
\begin{eqnarray*}
-c_{1}{r \over 2a} \sum_{x,k}\bar{\psi} (x) U(x,k) \psi (x+k)
\end{eqnarray*}
\begin{eqnarray}
-c_{2}{r \over 2a} \sum_{x,k}\bar{\psi} (x) U(x,2k) \psi (x+2k).
\label{ImprovedH}
\end{eqnarray}
Here $U(x,2k)=U(x,k)U(x+k,k)$ and  
$b_{1}, b_{2}, c_{1}$  and $c_{2}$ are chosen as 
\begin{eqnarray}
b_{1}={4 \over 3},  b_{2}=-{1 \over 6}, 
c_{1}={4 \over 3}, c_{2}=-{1 \over 3}
\end{eqnarray}
to classically cancel the $O(a)$ error.
These coefficients are the same for any d+1 dimensions
and gauge group. In 1+1D, there is only color-electric
energy in the gluonic sector, therefore
classical improvement is sufficient.

With the absence of the $O(ra)$ errors, we
expect that we can extract the continuum physics in a more reliable way.

\section{PHYSICAL RESULTS}

We have constructed the wave functions of vacuum and 
quark-antiquark vector state, 
and discussed the mixing problem of 
the operator $\langle \bar{\psi} \psi \rangle$
with the identity. For details, see Ref. \cite{Jiang}.

Figure 1 plots $-\langle \bar{\psi} \psi \rangle_{sub}/(gN_C)$ 
as a function of $1/g^2$ in SU(3) gauge theory ($N_C=3$)
for Wilson parameter $r=0.1$ (crosses) 
and $r=1$ (diamonds). Here $sub$ means the lattice data subtract 
the contribution of free fermions. 
Figure 3 plots  $aM_V/g$ as a function of $1/g^2$, with $M_V$
being the vector mass. 
As one sees, the results for $r=1$ deviate obviously from those 
for $r=0.1$, which is attributed to the $O(ra)$ error of the Wilson term.

The results from the improved Hamiltonian are presented in Fig. 2, 
and Fig. 4. One observes that the differences between the results 
for $r=1$ and $r=0$ are significantly reduced. 
Most impressively, the data for the quark condensate coincide each other.
A similar $r$ test has also been used by workers in Lagrangian
formulation for checking the efficiency of the improvement program.

For the interest of workers outside the lattice community \cite{tHooft74}, 
we also show the results as a function of $1/N_C^2$ in Fig. 5  
and Fig. 6.  

It is worth mentioning  the improved Hamiltonian formulation  
for pure gauge theory \cite{Luo98_1,Luo98_2} 
proposed by Luo, Guo, Kr\"oger, and Sch\"utte
is also giving promising results.

In conclusion, the results from the improved Hamiltonian \cite{Luo94}
have much less systematic errors than those from Wilson's.
We believe that its application to QCD in 3+1 dimensions will be very encouraging.

X.Q.L. is supported by the
National Natural Science Fund for Distinguished Young Scholars,
supplemented by the
National Natural Science Foundation of China, 
fund for international cooperation and exchange,
the Ministry of Education, 
the Zhongshan University Administrations
and Hong Kong Foundation of
the Zhongshan University Advanced Research Center. 
We also
thank the Lattice 98 organizers
for additional support and assistance.
H.K. would like to acknowledge support by NSERC Canada.

\begin{figure}[htb]
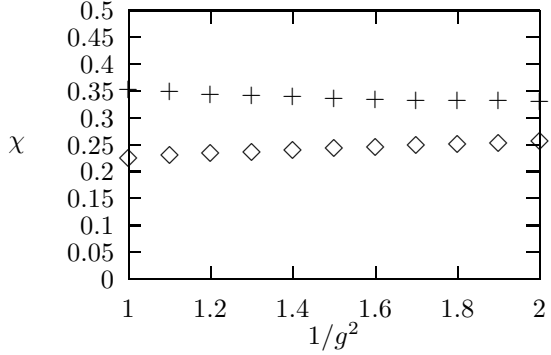

\hspace{6mm}
\setlength{\unitlength}{0.240900pt}
\ifx\plotpoint\undefined\newsavebox{\plotpoint}\fi
\sbox{\plotpoint}{\rule[-0.200pt]{0.400pt}{0.400pt}}%
\font\gnuplot=cmr10 at 10pt
\sbox{\plotpoint}{\rule[-0.200pt]{0.400pt}{0.400pt}}%
\put(220.0,113.0){\rule[-0.200pt]{155.621pt}{0.400pt}}
\put(220.0,113.0){\rule[-0.200pt]{4.818pt}{0.400pt}}
\put(198,113){\makebox(0,0)[r]{0}}
\put(846.0,113.0){\rule[-0.200pt]{4.818pt}{0.400pt}}
\put(220.0,155.0){\rule[-0.200pt]{4.818pt}{0.400pt}}
\put(198,155){\makebox(0,0)[r]{0.05}}
\put(846.0,155.0){\rule[-0.200pt]{4.818pt}{0.400pt}}
\put(220.0,197.0){\rule[-0.200pt]{4.818pt}{0.400pt}}
\put(198,197){\makebox(0,0)[r]{0.1}}
\put(846.0,197.0){\rule[-0.200pt]{4.818pt}{0.400pt}}
\put(220.0,240.0){\rule[-0.200pt]{4.818pt}{0.400pt}}
\put(198,240){\makebox(0,0)[r]{0.15}}
\put(846.0,240.0){\rule[-0.200pt]{4.818pt}{0.400pt}}
\put(220.0,282.0){\rule[-0.200pt]{4.818pt}{0.400pt}}
\put(198,282){\makebox(0,0)[r]{0.2}}
\put(846.0,282.0){\rule[-0.200pt]{4.818pt}{0.400pt}}
\put(220.0,324.0){\rule[-0.200pt]{4.818pt}{0.400pt}}
\put(198,324){\makebox(0,0)[r]{0.25}}
\put(846.0,324.0){\rule[-0.200pt]{4.818pt}{0.400pt}}
\put(220.0,366.0){\rule[-0.200pt]{4.818pt}{0.400pt}}
\put(198,366){\makebox(0,0)[r]{0.3}}
\put(846.0,366.0){\rule[-0.200pt]{4.818pt}{0.400pt}}
\put(220.0,408.0){\rule[-0.200pt]{4.818pt}{0.400pt}}
\put(198,408){\makebox(0,0)[r]{0.35}}
\put(846.0,408.0){\rule[-0.200pt]{4.818pt}{0.400pt}}
\put(220.0,451.0){\rule[-0.200pt]{4.818pt}{0.400pt}}
\put(198,451){\makebox(0,0)[r]{0.4}}
\put(846.0,451.0){\rule[-0.200pt]{4.818pt}{0.400pt}}
\put(220.0,493.0){\rule[-0.200pt]{4.818pt}{0.400pt}}
\put(198,493){\makebox(0,0)[r]{0.45}}
\put(846.0,493.0){\rule[-0.200pt]{4.818pt}{0.400pt}}
\put(220.0,535.0){\rule[-0.200pt]{4.818pt}{0.400pt}}
\put(198,535){\makebox(0,0)[r]{0.5}}
\put(846.0,535.0){\rule[-0.200pt]{4.818pt}{0.400pt}}
\put(220.0,113.0){\rule[-0.200pt]{0.400pt}{4.818pt}}
\put(220,68){\makebox(0,0){1}}
\put(220.0,515.0){\rule[-0.200pt]{0.400pt}{4.818pt}}
\put(349.0,113.0){\rule[-0.200pt]{0.400pt}{4.818pt}}
\put(349,68){\makebox(0,0){1.2}}
\put(349.0,515.0){\rule[-0.200pt]{0.400pt}{4.818pt}}
\put(478.0,113.0){\rule[-0.200pt]{0.400pt}{4.818pt}}
\put(478,68){\makebox(0,0){1.4}}
\put(478.0,515.0){\rule[-0.200pt]{0.400pt}{4.818pt}}
\put(608.0,113.0){\rule[-0.200pt]{0.400pt}{4.818pt}}
\put(608,68){\makebox(0,0){1.6}}
\put(608.0,515.0){\rule[-0.200pt]{0.400pt}{4.818pt}}
\put(737.0,113.0){\rule[-0.200pt]{0.400pt}{4.818pt}}
\put(737,68){\makebox(0,0){1.8}}
\put(737.0,515.0){\rule[-0.200pt]{0.400pt}{4.818pt}}
\put(866.0,113.0){\rule[-0.200pt]{0.400pt}{4.818pt}}
\put(866,68){\makebox(0,0){2}}
\put(866.0,515.0){\rule[-0.200pt]{0.400pt}{4.818pt}}
\put(220.0,113.0){\rule[-0.200pt]{155.621pt}{0.400pt}}
\put(866.0,113.0){\rule[-0.200pt]{0.400pt}{101.660pt}}
\put(220.0,535.0){\rule[-0.200pt]{155.621pt}{0.400pt}}
\put(45,324){\makebox(0,0){$\chi$}}
\put(543,23){\makebox(0,0){$1/g^2$}}
\put(220.0,113.0){\rule[-0.200pt]{0.400pt}{101.660pt}}
\put(220,302){\raisebox{-.8pt}{\makebox(0,0){$\Diamond$}}}
\put(285,306){\raisebox{-.8pt}{\makebox(0,0){$\Diamond$}}}
\put(349,309){\raisebox{-.8pt}{\makebox(0,0){$\Diamond$}}}
\put(414,312){\raisebox{-.8pt}{\makebox(0,0){$\Diamond$}}}
\put(478,314){\raisebox{-.8pt}{\makebox(0,0){$\Diamond$}}}
\put(543,317){\raisebox{-.8pt}{\makebox(0,0){$\Diamond$}}}
\put(608,319){\raisebox{-.8pt}{\makebox(0,0){$\Diamond$}}}
\put(672,322){\raisebox{-.8pt}{\makebox(0,0){$\Diamond$}}}
\put(737,324){\raisebox{-.8pt}{\makebox(0,0){$\Diamond$}}}
\put(801,326){\raisebox{-.8pt}{\makebox(0,0){$\Diamond$}}}
\put(866,328){\raisebox{-.8pt}{\makebox(0,0){$\Diamond$}}}
\put(220,412){\makebox(0,0){$+$}}
\put(285,408){\makebox(0,0){$+$}}
\put(349,404){\makebox(0,0){$+$}}
\put(414,402){\makebox(0,0){$+$}}
\put(478,400){\makebox(0,0){$+$}}
\put(543,397){\makebox(0,0){$+$}}
\put(608,396){\makebox(0,0){$+$}}
\put(672,395){\makebox(0,0){$+$}}
\put(737,394){\makebox(0,0){$+$}}
\put(801,394){\makebox(0,0){$+$}}
\put(866,393){\makebox(0,0){$+$}}
\vspace{-10mm}
\caption{$\chi=-\langle \bar{\psi} \psi \rangle_{sub}/(gN_c)$ versus $1/g^{2}$ 
for $N_C=3$ with Wilson fermions. 
Crosses: $r=0.1$, Diamonds: $r=1$.}
\label{fig1}
\end{figure}

\begin{figure}[htb]
\hspace{6mm}
\setlength{\unitlength}{0.240900pt}
\ifx\plotpoint\undefined\newsavebox{\plotpoint}\fi
\font\gnuplot=cmr10 at 10pt
\sbox{\plotpoint}{\rule[-0.200pt]{0.400pt}{0.400pt}}%
\put(220.0,113.0){\rule[-0.200pt]{155.621pt}{0.400pt}}
\put(220.0,113.0){\rule[-0.200pt]{4.818pt}{0.400pt}}
\put(198,113){\makebox(0,0)[r]{0}}
\put(846.0,113.0){\rule[-0.200pt]{4.818pt}{0.400pt}}
\put(220.0,155.0){\rule[-0.200pt]{4.818pt}{0.400pt}}
\put(198,155){\makebox(0,0)[r]{0.05}}
\put(846.0,155.0){\rule[-0.200pt]{4.818pt}{0.400pt}}
\put(220.0,197.0){\rule[-0.200pt]{4.818pt}{0.400pt}}
\put(198,197){\makebox(0,0)[r]{0.1}}
\put(846.0,197.0){\rule[-0.200pt]{4.818pt}{0.400pt}}
\put(220.0,240.0){\rule[-0.200pt]{4.818pt}{0.400pt}}
\put(198,240){\makebox(0,0)[r]{0.15}}
\put(846.0,240.0){\rule[-0.200pt]{4.818pt}{0.400pt}}
\put(220.0,282.0){\rule[-0.200pt]{4.818pt}{0.400pt}}
\put(198,282){\makebox(0,0)[r]{0.2}}
\put(846.0,282.0){\rule[-0.200pt]{4.818pt}{0.400pt}}
\put(220.0,324.0){\rule[-0.200pt]{4.818pt}{0.400pt}}
\put(198,324){\makebox(0,0)[r]{0.25}}
\put(846.0,324.0){\rule[-0.200pt]{4.818pt}{0.400pt}}
\put(220.0,366.0){\rule[-0.200pt]{4.818pt}{0.400pt}}
\put(198,366){\makebox(0,0)[r]{0.3}}
\put(846.0,366.0){\rule[-0.200pt]{4.818pt}{0.400pt}}
\put(220.0,408.0){\rule[-0.200pt]{4.818pt}{0.400pt}}
\put(198,408){\makebox(0,0)[r]{0.35}}
\put(846.0,408.0){\rule[-0.200pt]{4.818pt}{0.400pt}}
\put(220.0,451.0){\rule[-0.200pt]{4.818pt}{0.400pt}}
\put(198,451){\makebox(0,0)[r]{0.4}}
\put(846.0,451.0){\rule[-0.200pt]{4.818pt}{0.400pt}}
\put(220.0,493.0){\rule[-0.200pt]{4.818pt}{0.400pt}}
\put(198,493){\makebox(0,0)[r]{0.45}}
\put(846.0,493.0){\rule[-0.200pt]{4.818pt}{0.400pt}}
\put(220.0,535.0){\rule[-0.200pt]{4.818pt}{0.400pt}}
\put(198,535){\makebox(0,0)[r]{0.5}}
\put(846.0,535.0){\rule[-0.200pt]{4.818pt}{0.400pt}}
\put(220.0,113.0){\rule[-0.200pt]{0.400pt}{4.818pt}}
\put(220,68){\makebox(0,0){1}}
\put(220.0,515.0){\rule[-0.200pt]{0.400pt}{4.818pt}}
\put(349.0,113.0){\rule[-0.200pt]{0.400pt}{4.818pt}}
\put(349,68){\makebox(0,0){1.2}}
\put(349.0,515.0){\rule[-0.200pt]{0.400pt}{4.818pt}}
\put(478.0,113.0){\rule[-0.200pt]{0.400pt}{4.818pt}}
\put(478,68){\makebox(0,0){1.4}}
\put(478.0,515.0){\rule[-0.200pt]{0.400pt}{4.818pt}}
\put(608.0,113.0){\rule[-0.200pt]{0.400pt}{4.818pt}}
\put(608,68){\makebox(0,0){1.6}}
\put(608.0,515.0){\rule[-0.200pt]{0.400pt}{4.818pt}}
\put(737.0,113.0){\rule[-0.200pt]{0.400pt}{4.818pt}}
\put(737,68){\makebox(0,0){1.8}}
\put(737.0,515.0){\rule[-0.200pt]{0.400pt}{4.818pt}}
\put(866.0,113.0){\rule[-0.200pt]{0.400pt}{4.818pt}}
\put(866,68){\makebox(0,0){2}}
\put(866.0,515.0){\rule[-0.200pt]{0.400pt}{4.818pt}}
\put(220.0,113.0){\rule[-0.200pt]{155.621pt}{0.400pt}}
\put(866.0,113.0){\rule[-0.200pt]{0.400pt}{101.660pt}}
\put(220.0,535.0){\rule[-0.200pt]{155.621pt}{0.400pt}}
\put(45,324){\makebox(0,0){$\chi$}}
\put(543,23){\makebox(0,0){$1/g^2$}}
\put(220.0,113.0){\rule[-0.200pt]{0.400pt}{101.660pt}}
\put(220,340){\raisebox{-.8pt}{\makebox(0,0){$\Diamond$}}}
\put(285,342){\raisebox{-.8pt}{\makebox(0,0){$\Diamond$}}}
\put(349,344){\raisebox{-.8pt}{\makebox(0,0){$\Diamond$}}}
\put(414,344){\raisebox{-.8pt}{\makebox(0,0){$\Diamond$}}}
\put(478,345){\raisebox{-.8pt}{\makebox(0,0){$\Diamond$}}}
\put(543,347){\raisebox{-.8pt}{\makebox(0,0){$\Diamond$}}}
\put(608,347){\raisebox{-.8pt}{\makebox(0,0){$\Diamond$}}}
\put(672,348){\raisebox{-.8pt}{\makebox(0,0){$\Diamond$}}}
\put(737,349){\raisebox{-.8pt}{\makebox(0,0){$\Diamond$}}}
\put(801,350){\raisebox{-.8pt}{\makebox(0,0){$\Diamond$}}}
\put(866,352){\raisebox{-.8pt}{\makebox(0,0){$\Diamond$}}}
\put(220,357){\makebox(0,0){$+$}}
\put(285,356){\makebox(0,0){$+$}}
\put(349,356){\makebox(0,0){$+$}}
\put(414,355){\makebox(0,0){$+$}}
\put(478,354){\makebox(0,0){$+$}}
\put(543,353){\makebox(0,0){$+$}}
\put(608,352){\makebox(0,0){$+$}}
\put(672,352){\makebox(0,0){$+$}}
\put(737,352){\makebox(0,0){$+$}}
\put(801,352){\makebox(0,0){$+$}}
\put(866,352){\makebox(0,0){$+$}}
\vspace{-10mm}
\caption{$\chi=-\langle \bar{\psi} \psi \rangle_{sub}/(gN_c)$ versus $1/g^{2}$ 
for $N_C=3$ with improved Wilson fermions. 
Crosses: $r=0.1$, Diamonds: $r=1$.}
\label{fig2}
\end{figure}

\begin{figure}[htb]
\hspace{6mm}
\setlength{\unitlength}{0.240900pt}
\ifx\plotpoint\undefined\newsavebox{\plotpoint}\fi
\font\gnuplot=cmr10 at 10pt
\sbox{\plotpoint}{\rule[-0.200pt]{0.400pt}{0.400pt}}%
\put(220.0,113.0){\rule[-0.200pt]{155.621pt}{0.400pt}}
\put(220.0,113.0){\rule[-0.200pt]{4.818pt}{0.400pt}}
\put(198,113){\makebox(0,0)[r]{0}}
\put(846.0,113.0){\rule[-0.200pt]{4.818pt}{0.400pt}}
\put(220.0,166.0){\rule[-0.200pt]{4.818pt}{0.400pt}}
\put(198,166){\makebox(0,0)[r]{0.2}}
\put(846.0,166.0){\rule[-0.200pt]{4.818pt}{0.400pt}}
\put(220.0,219.0){\rule[-0.200pt]{4.818pt}{0.400pt}}
\put(198,219){\makebox(0,0)[r]{0.4}}
\put(846.0,219.0){\rule[-0.200pt]{4.818pt}{0.400pt}}
\put(220.0,271.0){\rule[-0.200pt]{4.818pt}{0.400pt}}
\put(198,271){\makebox(0,0)[r]{0.6}}
\put(846.0,271.0){\rule[-0.200pt]{4.818pt}{0.400pt}}
\put(220.0,324.0){\rule[-0.200pt]{4.818pt}{0.400pt}}
\put(198,324){\makebox(0,0)[r]{0.8}}
\put(846.0,324.0){\rule[-0.200pt]{4.818pt}{0.400pt}}
\put(220.0,377.0){\rule[-0.200pt]{4.818pt}{0.400pt}}
\put(198,377){\makebox(0,0)[r]{1}}
\put(846.0,377.0){\rule[-0.200pt]{4.818pt}{0.400pt}}
\put(220.0,429.0){\rule[-0.200pt]{4.818pt}{0.400pt}}
\put(198,429){\makebox(0,0)[r]{1.2}}
\put(846.0,429.0){\rule[-0.200pt]{4.818pt}{0.400pt}}
\put(220.0,482.0){\rule[-0.200pt]{4.818pt}{0.400pt}}
\put(198,482){\makebox(0,0)[r]{1.4}}
\put(846.0,482.0){\rule[-0.200pt]{4.818pt}{0.400pt}}
\put(220.0,535.0){\rule[-0.200pt]{4.818pt}{0.400pt}}
\put(198,535){\makebox(0,0)[r]{1.6}}
\put(846.0,535.0){\rule[-0.200pt]{4.818pt}{0.400pt}}
\put(220.0,113.0){\rule[-0.200pt]{0.400pt}{4.818pt}}
\put(220,68){\makebox(0,0){1}}
\put(220.0,515.0){\rule[-0.200pt]{0.400pt}{4.818pt}}
\put(349.0,113.0){\rule[-0.200pt]{0.400pt}{4.818pt}}
\put(349,68){\makebox(0,0){1.2}}
\put(349.0,515.0){\rule[-0.200pt]{0.400pt}{4.818pt}}
\put(478.0,113.0){\rule[-0.200pt]{0.400pt}{4.818pt}}
\put(478,68){\makebox(0,0){1.4}}
\put(478.0,515.0){\rule[-0.200pt]{0.400pt}{4.818pt}}
\put(608.0,113.0){\rule[-0.200pt]{0.400pt}{4.818pt}}
\put(608,68){\makebox(0,0){1.6}}
\put(608.0,515.0){\rule[-0.200pt]{0.400pt}{4.818pt}}
\put(737.0,113.0){\rule[-0.200pt]{0.400pt}{4.818pt}}
\put(737,68){\makebox(0,0){1.8}}
\put(737.0,515.0){\rule[-0.200pt]{0.400pt}{4.818pt}}
\put(866.0,113.0){\rule[-0.200pt]{0.400pt}{4.818pt}}
\put(866,68){\makebox(0,0){2}}
\put(866.0,515.0){\rule[-0.200pt]{0.400pt}{4.818pt}}
\put(220.0,113.0){\rule[-0.200pt]{155.621pt}{0.400pt}}
\put(866.0,113.0){\rule[-0.200pt]{0.400pt}{101.660pt}}
\put(220.0,535.0){\rule[-0.200pt]{155.621pt}{0.400pt}}
\put(45,324){\makebox(0,0){$aM_V/g$}}
\put(543,23){\makebox(0,0){$1/g^2$}}
\put(220.0,113.0){\rule[-0.200pt]{0.400pt}{101.660pt}}
\put(220,366){\raisebox{-.8pt}{\makebox(0,0){$\Diamond$}}}
\put(349,371){\raisebox{-.8pt}{\makebox(0,0){$\Diamond$}}}
\put(478,376){\raisebox{-.8pt}{\makebox(0,0){$\Diamond$}}}
\put(608,381){\raisebox{-.8pt}{\makebox(0,0){$\Diamond$}}}
\put(737,386){\raisebox{-.8pt}{\makebox(0,0){$\Diamond$}}}
\put(866,391){\raisebox{-.8pt}{\makebox(0,0){$\Diamond$}}}
\put(220,264){\makebox(0,0){$+$}}
\put(349,278){\makebox(0,0){$+$}}
\put(478,289){\makebox(0,0){$+$}}
\put(608,298){\makebox(0,0){$+$}}
\put(737,304){\makebox(0,0){$+$}}
\put(866,309){\makebox(0,0){$+$}}
\vspace{-10mm}
\caption{$aM_{V}/g$ versus $1/g^{2}$ 
for $N_C=3$ with Wilson fermions. 
Crosses: $r=0.1$, Diamonds: $r=1$.}
\label{fig3}
\end{figure}

\begin{figure}[htb]
\hspace{6mm}
\setlength{\unitlength}{0.240900pt}
\ifx\plotpoint\undefined\newsavebox{\plotpoint}\fi
\font\gnuplot=cmr10 at 10pt
\sbox{\plotpoint}{\rule[-0.200pt]{0.400pt}{0.400pt}}%
\put(220.0,113.0){\rule[-0.200pt]{155.621pt}{0.400pt}}
\put(220.0,113.0){\rule[-0.200pt]{4.818pt}{0.400pt}}
\put(198,113){\makebox(0,0)[r]{0}}
\put(846.0,113.0){\rule[-0.200pt]{4.818pt}{0.400pt}}
\put(220.0,166.0){\rule[-0.200pt]{4.818pt}{0.400pt}}
\put(198,166){\makebox(0,0)[r]{0.2}}
\put(846.0,166.0){\rule[-0.200pt]{4.818pt}{0.400pt}}
\put(220.0,219.0){\rule[-0.200pt]{4.818pt}{0.400pt}}
\put(198,219){\makebox(0,0)[r]{0.4}}
\put(846.0,219.0){\rule[-0.200pt]{4.818pt}{0.400pt}}
\put(220.0,271.0){\rule[-0.200pt]{4.818pt}{0.400pt}}
\put(198,271){\makebox(0,0)[r]{0.6}}
\put(846.0,271.0){\rule[-0.200pt]{4.818pt}{0.400pt}}
\put(220.0,324.0){\rule[-0.200pt]{4.818pt}{0.400pt}}
\put(198,324){\makebox(0,0)[r]{0.8}}
\put(846.0,324.0){\rule[-0.200pt]{4.818pt}{0.400pt}}
\put(220.0,377.0){\rule[-0.200pt]{4.818pt}{0.400pt}}
\put(198,377){\makebox(0,0)[r]{1}}
\put(846.0,377.0){\rule[-0.200pt]{4.818pt}{0.400pt}}
\put(220.0,429.0){\rule[-0.200pt]{4.818pt}{0.400pt}}
\put(198,429){\makebox(0,0)[r]{1.2}}
\put(846.0,429.0){\rule[-0.200pt]{4.818pt}{0.400pt}}
\put(220.0,482.0){\rule[-0.200pt]{4.818pt}{0.400pt}}
\put(198,482){\makebox(0,0)[r]{1.4}}
\put(846.0,482.0){\rule[-0.200pt]{4.818pt}{0.400pt}}
\put(220.0,535.0){\rule[-0.200pt]{4.818pt}{0.400pt}}
\put(198,535){\makebox(0,0)[r]{1.6}}
\put(846.0,535.0){\rule[-0.200pt]{4.818pt}{0.400pt}}
\put(220.0,113.0){\rule[-0.200pt]{0.400pt}{4.818pt}}
\put(220,68){\makebox(0,0){1}}
\put(220.0,515.0){\rule[-0.200pt]{0.400pt}{4.818pt}}
\put(349.0,113.0){\rule[-0.200pt]{0.400pt}{4.818pt}}
\put(349,68){\makebox(0,0){1.2}}
\put(349.0,515.0){\rule[-0.200pt]{0.400pt}{4.818pt}}
\put(478.0,113.0){\rule[-0.200pt]{0.400pt}{4.818pt}}
\put(478,68){\makebox(0,0){1.4}}
\put(478.0,515.0){\rule[-0.200pt]{0.400pt}{4.818pt}}
\put(608.0,113.0){\rule[-0.200pt]{0.400pt}{4.818pt}}
\put(608,68){\makebox(0,0){1.6}}
\put(608.0,515.0){\rule[-0.200pt]{0.400pt}{4.818pt}}
\put(737.0,113.0){\rule[-0.200pt]{0.400pt}{4.818pt}}
\put(737,68){\makebox(0,0){1.8}}
\put(737.0,515.0){\rule[-0.200pt]{0.400pt}{4.818pt}}
\put(866.0,113.0){\rule[-0.200pt]{0.400pt}{4.818pt}}
\put(866,68){\makebox(0,0){2}}
\put(866.0,515.0){\rule[-0.200pt]{0.400pt}{4.818pt}}
\put(220.0,113.0){\rule[-0.200pt]{155.621pt}{0.400pt}}
\put(866.0,113.0){\rule[-0.200pt]{0.400pt}{101.660pt}}
\put(220.0,535.0){\rule[-0.200pt]{155.621pt}{0.400pt}}
\put(45,324){\makebox(0,0){$aM_V/g$}}
\put(543,23){\makebox(0,0){$1/g^2$}}
\put(220.0,113.0){\rule[-0.200pt]{0.400pt}{101.660pt}}
\put(220,332){\raisebox{-.8pt}{\makebox(0,0){$\Diamond$}}}
\put(349,334){\raisebox{-.8pt}{\makebox(0,0){$\Diamond$}}}
\put(478,335){\raisebox{-.8pt}{\makebox(0,0){$\Diamond$}}}
\put(608,337){\raisebox{-.8pt}{\makebox(0,0){$\Diamond$}}}
\put(737,339){\raisebox{-.8pt}{\makebox(0,0){$\Diamond$}}}
\put(866,341){\raisebox{-.8pt}{\makebox(0,0){$\Diamond$}}}
\put(220,293){\makebox(0,0){$+$}}
\put(349,303){\makebox(0,0){$+$}}
\put(478,308){\makebox(0,0){$+$}}
\put(608,311){\makebox(0,0){$+$}}
\put(737,313){\makebox(0,0){$+$}}
\put(866,315){\makebox(0,0){$+$}}
\vspace{-10mm}
\caption{$aM_{V}/g$ versus $1/g^{2}$ 
for $N_C=3$ with improved Wilson fermions. 
Crosses: $r=0.1$, Diamonds: $r=1$.}
\label{fig4}
\end{figure}

\begin{figure}[htb]
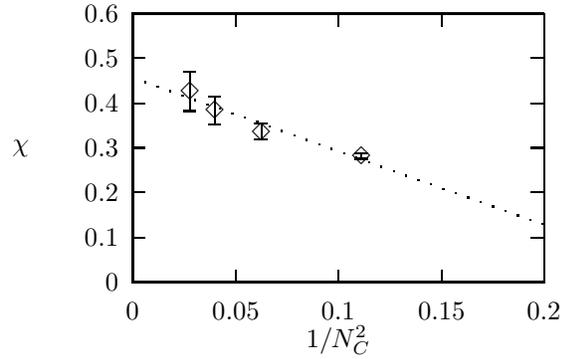

\hspace{6mm}
\setlength{\unitlength}{0.240900pt}
\ifx\plotpoint\undefined\newsavebox{\plotpoint}\fi
\font\gnuplot=cmr10 at 10pt
\sbox{\plotpoint}{\rule[-0.200pt]{0.400pt}{0.400pt}}%
\put(220.0,113.0){\rule[-0.200pt]{155.621pt}{0.400pt}}
\put(220.0,113.0){\rule[-0.200pt]{0.400pt}{101.660pt}}
\put(220.0,113.0){\rule[-0.200pt]{4.818pt}{0.400pt}}
\put(198,113){\makebox(0,0)[r]{0}}
\put(846.0,113.0){\rule[-0.200pt]{4.818pt}{0.400pt}}
\put(220.0,183.0){\rule[-0.200pt]{4.818pt}{0.400pt}}
\put(198,183){\makebox(0,0)[r]{0.1}}
\put(846.0,183.0){\rule[-0.200pt]{4.818pt}{0.400pt}}
\put(220.0,254.0){\rule[-0.200pt]{4.818pt}{0.400pt}}
\put(198,254){\makebox(0,0)[r]{0.2}}
\put(846.0,254.0){\rule[-0.200pt]{4.818pt}{0.400pt}}
\put(220.0,324.0){\rule[-0.200pt]{4.818pt}{0.400pt}}
\put(198,324){\makebox(0,0)[r]{0.3}}
\put(846.0,324.0){\rule[-0.200pt]{4.818pt}{0.400pt}}
\put(220.0,394.0){\rule[-0.200pt]{4.818pt}{0.400pt}}
\put(198,394){\makebox(0,0)[r]{0.4}}
\put(846.0,394.0){\rule[-0.200pt]{4.818pt}{0.400pt}}
\put(220.0,465.0){\rule[-0.200pt]{4.818pt}{0.400pt}}
\put(198,465){\makebox(0,0)[r]{0.5}}
\put(846.0,465.0){\rule[-0.200pt]{4.818pt}{0.400pt}}
\put(220.0,535.0){\rule[-0.200pt]{4.818pt}{0.400pt}}
\put(198,535){\makebox(0,0)[r]{0.6}}
\put(846.0,535.0){\rule[-0.200pt]{4.818pt}{0.400pt}}
\put(220.0,113.0){\rule[-0.200pt]{0.400pt}{4.818pt}}
\put(220,68){\makebox(0,0){0}}
\put(220.0,515.0){\rule[-0.200pt]{0.400pt}{4.818pt}}
\put(382.0,113.0){\rule[-0.200pt]{0.400pt}{4.818pt}}
\put(382,68){\makebox(0,0){0.05}}
\put(382.0,515.0){\rule[-0.200pt]{0.400pt}{4.818pt}}
\put(543.0,113.0){\rule[-0.200pt]{0.400pt}{4.818pt}}
\put(543,68){\makebox(0,0){0.1}}
\put(543.0,515.0){\rule[-0.200pt]{0.400pt}{4.818pt}}
\put(705.0,113.0){\rule[-0.200pt]{0.400pt}{4.818pt}}
\put(705,68){\makebox(0,0){0.15}}
\put(705.0,515.0){\rule[-0.200pt]{0.400pt}{4.818pt}}
\put(866.0,113.0){\rule[-0.200pt]{0.400pt}{4.818pt}}
\put(866,68){\makebox(0,0){0.2}}
\put(866.0,515.0){\rule[-0.200pt]{0.400pt}{4.818pt}}
\put(220.0,113.0){\rule[-0.200pt]{155.621pt}{0.400pt}}
\put(866.0,113.0){\rule[-0.200pt]{0.400pt}{101.660pt}}
\put(220.0,535.0){\rule[-0.200pt]{155.621pt}{0.400pt}}
\put(45,324){\makebox(0,0){$\chi$}}
\put(543,23){\makebox(0,0){$1/N_C^2$}}
\put(220.0,113.0){\rule[-0.200pt]{0.400pt}{101.660pt}}
\put(579,311){\raisebox{-.8pt}{\makebox(0,0){$\Diamond$}}}
\put(422,349){\raisebox{-.8pt}{\makebox(0,0){$\Diamond$}}}
\put(349,383){\raisebox{-.8pt}{\makebox(0,0){$\Diamond$}}}
\put(310,413){\raisebox{-.8pt}{\makebox(0,0){$\Diamond$}}}
\put(579.0,307.0){\rule[-0.200pt]{0.400pt}{2.168pt}}
\put(569.0,307.0){\rule[-0.200pt]{4.818pt}{0.400pt}}
\put(569.0,316.0){\rule[-0.200pt]{4.818pt}{0.400pt}}
\put(422.0,337.0){\rule[-0.200pt]{0.400pt}{6.022pt}}
\put(412.0,337.0){\rule[-0.200pt]{4.818pt}{0.400pt}}
\put(412.0,362.0){\rule[-0.200pt]{4.818pt}{0.400pt}}
\put(349.0,361.0){\rule[-0.200pt]{0.400pt}{10.600pt}}
\put(339.0,361.0){\rule[-0.200pt]{4.818pt}{0.400pt}}
\put(339.0,405.0){\rule[-0.200pt]{4.818pt}{0.400pt}}
\put(310.0,382.0){\rule[-0.200pt]{0.400pt}{14.936pt}}
\put(300.0,382.0){\rule[-0.200pt]{4.818pt}{0.400pt}}
\put(300.0,444.0){\rule[-0.200pt]{4.818pt}{0.400pt}}
\put(220,434){\usebox{\plotpoint}}
\put(220.00,434.00){\usebox{\plotpoint}}
\multiput(227,431)(19.690,-6.563){0}{\usebox{\plotpoint}}
\put(239.55,427.13){\usebox{\plotpoint}}
\multiput(240,427)(18.564,-9.282){0}{\usebox{\plotpoint}}
\multiput(246,424)(19.957,-5.702){0}{\usebox{\plotpoint}}
\put(258.98,420.01){\usebox{\plotpoint}}
\multiput(259,420)(19.077,-8.176){0}{\usebox{\plotpoint}}
\multiput(266,417)(19.690,-6.563){0}{\usebox{\plotpoint}}
\put(278.53,413.13){\usebox{\plotpoint}}
\multiput(279,413)(18.564,-9.282){0}{\usebox{\plotpoint}}
\multiput(285,410)(19.957,-5.702){0}{\usebox{\plotpoint}}
\put(297.96,406.01){\usebox{\plotpoint}}
\multiput(298,406)(19.077,-8.176){0}{\usebox{\plotpoint}}
\multiput(305,403)(19.690,-6.563){0}{\usebox{\plotpoint}}
\put(317.51,399.14){\usebox{\plotpoint}}
\multiput(318,399)(18.564,-9.282){0}{\usebox{\plotpoint}}
\multiput(324,396)(19.957,-5.702){0}{\usebox{\plotpoint}}
\put(336.94,392.02){\usebox{\plotpoint}}
\multiput(337,392)(19.077,-8.176){0}{\usebox{\plotpoint}}
\multiput(344,389)(19.957,-5.702){0}{\usebox{\plotpoint}}
\put(356.50,385.17){\usebox{\plotpoint}}
\multiput(357,385)(19.077,-8.176){0}{\usebox{\plotpoint}}
\multiput(364,382)(19.690,-6.563){0}{\usebox{\plotpoint}}
\put(376.04,378.27){\usebox{\plotpoint}}
\multiput(377,378)(18.564,-9.282){0}{\usebox{\plotpoint}}
\multiput(383,375)(19.957,-5.702){0}{\usebox{\plotpoint}}
\put(395.47,371.18){\usebox{\plotpoint}}
\multiput(396,371)(19.077,-8.176){0}{\usebox{\plotpoint}}
\multiput(403,368)(19.690,-6.563){0}{\usebox{\plotpoint}}
\put(415.02,364.28){\usebox{\plotpoint}}
\multiput(416,364)(18.564,-9.282){0}{\usebox{\plotpoint}}
\multiput(422,361)(19.957,-5.702){0}{\usebox{\plotpoint}}
\put(434.45,357.18){\usebox{\plotpoint}}
\multiput(435,357)(19.077,-8.176){0}{\usebox{\plotpoint}}
\multiput(442,354)(19.690,-6.563){0}{\usebox{\plotpoint}}
\put(454.00,350.29){\usebox{\plotpoint}}
\multiput(455,350)(18.564,-9.282){0}{\usebox{\plotpoint}}
\multiput(461,347)(19.957,-5.702){0}{\usebox{\plotpoint}}
\put(473.43,343.19){\usebox{\plotpoint}}
\multiput(474,343)(19.077,-8.176){0}{\usebox{\plotpoint}}
\multiput(481,340)(19.957,-5.702){0}{\usebox{\plotpoint}}
\put(492.99,336.34){\usebox{\plotpoint}}
\multiput(494,336)(19.077,-8.176){0}{\usebox{\plotpoint}}
\multiput(501,333)(19.690,-6.563){0}{\usebox{\plotpoint}}
\put(512.53,329.42){\usebox{\plotpoint}}
\multiput(514,329)(18.564,-9.282){0}{\usebox{\plotpoint}}
\multiput(520,326)(19.957,-5.702){0}{\usebox{\plotpoint}}
\put(531.97,322.34){\usebox{\plotpoint}}
\multiput(533,322)(19.957,-5.702){0}{\usebox{\plotpoint}}
\multiput(540,320)(18.564,-9.282){0}{\usebox{\plotpoint}}
\put(551.46,315.44){\usebox{\plotpoint}}
\multiput(553,315)(19.690,-6.563){0}{\usebox{\plotpoint}}
\multiput(559,313)(19.077,-8.176){0}{\usebox{\plotpoint}}
\put(570.95,308.35){\usebox{\plotpoint}}
\multiput(572,308)(19.957,-5.702){0}{\usebox{\plotpoint}}
\multiput(579,306)(18.564,-9.282){0}{\usebox{\plotpoint}}
\put(590.44,301.45){\usebox{\plotpoint}}
\multiput(592,301)(19.690,-6.563){0}{\usebox{\plotpoint}}
\multiput(598,299)(19.077,-8.176){0}{\usebox{\plotpoint}}
\put(609.99,294.57){\usebox{\plotpoint}}
\multiput(612,294)(19.690,-6.563){0}{\usebox{\plotpoint}}
\multiput(618,292)(19.077,-8.176){0}{\usebox{\plotpoint}}
\put(629.49,287.50){\usebox{\plotpoint}}
\multiput(631,287)(19.957,-5.702){0}{\usebox{\plotpoint}}
\multiput(638,285)(18.564,-9.282){0}{\usebox{\plotpoint}}
\put(648.97,280.58){\usebox{\plotpoint}}
\multiput(651,280)(19.690,-6.563){0}{\usebox{\plotpoint}}
\multiput(657,278)(19.077,-8.176){0}{\usebox{\plotpoint}}
\put(668.47,273.51){\usebox{\plotpoint}}
\multiput(670,273)(19.957,-5.702){0}{\usebox{\plotpoint}}
\multiput(677,271)(18.564,-9.282){0}{\usebox{\plotpoint}}
\put(687.95,266.59){\usebox{\plotpoint}}
\multiput(690,266)(19.690,-6.563){0}{\usebox{\plotpoint}}
\multiput(696,264)(19.077,-8.176){0}{\usebox{\plotpoint}}
\put(707.44,259.52){\usebox{\plotpoint}}
\multiput(709,259)(19.957,-5.702){0}{\usebox{\plotpoint}}
\multiput(716,257)(18.564,-9.282){0}{\usebox{\plotpoint}}
\put(726.93,252.59){\usebox{\plotpoint}}
\multiput(729,252)(19.690,-6.563){0}{\usebox{\plotpoint}}
\multiput(735,250)(19.077,-8.176){0}{\usebox{\plotpoint}}
\put(746.48,245.72){\usebox{\plotpoint}}
\multiput(749,245)(19.690,-6.563){0}{\usebox{\plotpoint}}
\multiput(755,243)(19.077,-8.176){0}{\usebox{\plotpoint}}
\put(765.98,238.67){\usebox{\plotpoint}}
\multiput(768,238)(19.957,-5.702){0}{\usebox{\plotpoint}}
\multiput(775,236)(18.564,-9.282){0}{\usebox{\plotpoint}}
\put(785.46,231.73){\usebox{\plotpoint}}
\multiput(788,231)(19.690,-6.563){0}{\usebox{\plotpoint}}
\multiput(794,229)(19.077,-8.176){0}{\usebox{\plotpoint}}
\put(804.96,224.68){\usebox{\plotpoint}}
\multiput(807,224)(19.957,-5.702){0}{\usebox{\plotpoint}}
\multiput(814,222)(18.564,-9.282){0}{\usebox{\plotpoint}}
\put(824.44,217.73){\usebox{\plotpoint}}
\multiput(827,217)(19.690,-6.563){0}{\usebox{\plotpoint}}
\multiput(833,215)(19.077,-8.176){0}{\usebox{\plotpoint}}
\put(843.94,210.69){\usebox{\plotpoint}}
\multiput(846,210)(19.957,-5.702){0}{\usebox{\plotpoint}}
\multiput(853,208)(18.564,-9.282){0}{\usebox{\plotpoint}}
\put(863.42,203.74){\usebox{\plotpoint}}
\put(866,203){\usebox{\plotpoint}}
\vspace{-10mm}
\caption{$\chi=-\langle \bar{\psi} \psi \rangle_{cont}/(eN_C)$ 
versus $1/N_C^{2}$ in the continuum.
The error bars are estimated  from the data for
different $r$.}
\label{fig5}
\end{figure}

\begin{figure}[htb]
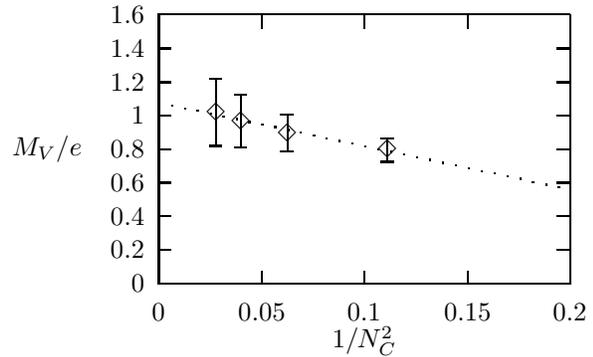

\hspace{6mm}
\setlength{\unitlength}{0.240900pt}
\ifx\plotpoint\undefined\newsavebox{\plotpoint}\fi
\sbox{\plotpoint}{\rule[-0.200pt]{0.400pt}{0.400pt}}%
\font\gnuplot=cmr10 at 10pt
\sbox{\plotpoint}{\rule[-0.200pt]{0.400pt}{0.400pt}}%
\put(220.0,113.0){\rule[-0.200pt]{155.621pt}{0.400pt}}
\put(220.0,113.0){\rule[-0.200pt]{0.400pt}{101.660pt}}
\put(220.0,113.0){\rule[-0.200pt]{4.818pt}{0.400pt}}
\put(198,113){\makebox(0,0)[r]{0}}
\put(846.0,113.0){\rule[-0.200pt]{4.818pt}{0.400pt}}
\put(220.0,166.0){\rule[-0.200pt]{4.818pt}{0.400pt}}
\put(198,166){\makebox(0,0)[r]{0.2}}
\put(846.0,166.0){\rule[-0.200pt]{4.818pt}{0.400pt}}
\put(220.0,219.0){\rule[-0.200pt]{4.818pt}{0.400pt}}
\put(198,219){\makebox(0,0)[r]{0.4}}
\put(846.0,219.0){\rule[-0.200pt]{4.818pt}{0.400pt}}
\put(220.0,271.0){\rule[-0.200pt]{4.818pt}{0.400pt}}
\put(198,271){\makebox(0,0)[r]{0.6}}
\put(846.0,271.0){\rule[-0.200pt]{4.818pt}{0.400pt}}
\put(220.0,324.0){\rule[-0.200pt]{4.818pt}{0.400pt}}
\put(198,324){\makebox(0,0)[r]{0.8}}
\put(846.0,324.0){\rule[-0.200pt]{4.818pt}{0.400pt}}
\put(220.0,377.0){\rule[-0.200pt]{4.818pt}{0.400pt}}
\put(198,377){\makebox(0,0)[r]{1}}
\put(846.0,377.0){\rule[-0.200pt]{4.818pt}{0.400pt}}
\put(220.0,429.0){\rule[-0.200pt]{4.818pt}{0.400pt}}
\put(198,429){\makebox(0,0)[r]{1.2}}
\put(846.0,429.0){\rule[-0.200pt]{4.818pt}{0.400pt}}
\put(220.0,482.0){\rule[-0.200pt]{4.818pt}{0.400pt}}
\put(198,482){\makebox(0,0)[r]{1.4}}
\put(846.0,482.0){\rule[-0.200pt]{4.818pt}{0.400pt}}
\put(220.0,535.0){\rule[-0.200pt]{4.818pt}{0.400pt}}
\put(198,535){\makebox(0,0)[r]{1.6}}
\put(846.0,535.0){\rule[-0.200pt]{4.818pt}{0.400pt}}
\put(220.0,113.0){\rule[-0.200pt]{0.400pt}{4.818pt}}
\put(220,68){\makebox(0,0){0}}
\put(220.0,515.0){\rule[-0.200pt]{0.400pt}{4.818pt}}
\put(382.0,113.0){\rule[-0.200pt]{0.400pt}{4.818pt}}
\put(382,68){\makebox(0,0){0.05}}
\put(382.0,515.0){\rule[-0.200pt]{0.400pt}{4.818pt}}
\put(543.0,113.0){\rule[-0.200pt]{0.400pt}{4.818pt}}
\put(543,68){\makebox(0,0){0.1}}
\put(543.0,515.0){\rule[-0.200pt]{0.400pt}{4.818pt}}
\put(705.0,113.0){\rule[-0.200pt]{0.400pt}{4.818pt}}
\put(705,68){\makebox(0,0){0.15}}
\put(705.0,515.0){\rule[-0.200pt]{0.400pt}{4.818pt}}
\put(866.0,113.0){\rule[-0.200pt]{0.400pt}{4.818pt}}
\put(866,68){\makebox(0,0){0.2}}
\put(866.0,515.0){\rule[-0.200pt]{0.400pt}{4.818pt}}
\put(220.0,113.0){\rule[-0.200pt]{155.621pt}{0.400pt}}
\put(866.0,113.0){\rule[-0.200pt]{0.400pt}{101.660pt}}
\put(220.0,535.0){\rule[-0.200pt]{155.621pt}{0.400pt}}
\put(45,324){\makebox(0,0){$M_V/e$}}
\put(543,23){\makebox(0,0){$1/N_C^2$}}
\put(220.0,113.0){\rule[-0.200pt]{0.400pt}{101.660pt}}
\put(579,323){\raisebox{-.8pt}{\makebox(0,0){$\Diamond$}}}
\put(422,349){\raisebox{-.8pt}{\makebox(0,0){$\Diamond$}}}
\put(349,368){\raisebox{-.8pt}{\makebox(0,0){$\Diamond$}}}
\put(310,381){\raisebox{-.8pt}{\makebox(0,0){$\Diamond$}}}
\put(579.0,304.0){\rule[-0.200pt]{0.400pt}{8.913pt}}
\put(569.0,304.0){\rule[-0.200pt]{4.818pt}{0.400pt}}
\put(569.0,341.0){\rule[-0.200pt]{4.818pt}{0.400pt}}
\put(422.0,320.0){\rule[-0.200pt]{0.400pt}{13.972pt}}
\put(412.0,320.0){\rule[-0.200pt]{4.818pt}{0.400pt}}
\put(412.0,378.0){\rule[-0.200pt]{4.818pt}{0.400pt}}
\put(349.0,327.0){\rule[-0.200pt]{0.400pt}{19.754pt}}
\put(339.0,327.0){\rule[-0.200pt]{4.818pt}{0.400pt}}
\put(339.0,409.0){\rule[-0.200pt]{4.818pt}{0.400pt}}
\put(310.0,329.0){\rule[-0.200pt]{0.400pt}{25.294pt}}
\put(300.0,329.0){\rule[-0.200pt]{4.818pt}{0.400pt}}
\put(300.0,434.0){\rule[-0.200pt]{4.818pt}{0.400pt}}
\put(220,396){\usebox{\plotpoint}}
\put(220.00,396.00){\usebox{\plotpoint}}
\multiput(227,395)(20.473,-3.412){0}{\usebox{\plotpoint}}
\multiput(233,394)(19.957,-5.702){0}{\usebox{\plotpoint}}
\put(240.32,391.95){\usebox{\plotpoint}}
\multiput(246,391)(19.957,-5.702){0}{\usebox{\plotpoint}}
\multiput(253,389)(20.473,-3.412){0}{\usebox{\plotpoint}}
\put(260.62,387.77){\usebox{\plotpoint}}
\multiput(266,387)(19.690,-6.563){0}{\usebox{\plotpoint}}
\multiput(272,385)(20.547,-2.935){0}{\usebox{\plotpoint}}
\put(280.89,383.68){\usebox{\plotpoint}}
\multiput(285,383)(19.957,-5.702){0}{\usebox{\plotpoint}}
\multiput(292,381)(20.473,-3.412){0}{\usebox{\plotpoint}}
\put(301.11,379.11){\usebox{\plotpoint}}
\multiput(305,378)(20.473,-3.412){0}{\usebox{\plotpoint}}
\multiput(311,377)(20.547,-2.935){0}{\usebox{\plotpoint}}
\put(321.37,374.88){\usebox{\plotpoint}}
\multiput(324,374)(20.547,-2.935){0}{\usebox{\plotpoint}}
\multiput(331,373)(20.473,-3.412){0}{\usebox{\plotpoint}}
\put(341.64,370.67){\usebox{\plotpoint}}
\multiput(344,370)(20.547,-2.935){0}{\usebox{\plotpoint}}
\multiput(351,369)(19.690,-6.563){0}{\usebox{\plotpoint}}
\put(361.86,366.31){\usebox{\plotpoint}}
\multiput(364,366)(20.473,-3.412){0}{\usebox{\plotpoint}}
\multiput(370,365)(19.957,-5.702){0}{\usebox{\plotpoint}}
\put(382.16,362.14){\usebox{\plotpoint}}
\multiput(383,362)(20.547,-2.935){0}{\usebox{\plotpoint}}
\multiput(390,361)(19.690,-6.563){0}{\usebox{\plotpoint}}
\put(402.44,358.08){\usebox{\plotpoint}}
\multiput(403,358)(19.690,-6.563){0}{\usebox{\plotpoint}}
\multiput(409,356)(20.547,-2.935){0}{\usebox{\plotpoint}}
\multiput(416,355)(20.473,-3.412){0}{\usebox{\plotpoint}}
\put(422.69,353.80){\usebox{\plotpoint}}
\multiput(429,352)(20.473,-3.412){0}{\usebox{\plotpoint}}
\multiput(435,351)(20.547,-2.935){0}{\usebox{\plotpoint}}
\put(442.98,349.67){\usebox{\plotpoint}}
\multiput(448,348)(20.547,-2.935){0}{\usebox{\plotpoint}}
\multiput(455,347)(19.690,-6.563){0}{\usebox{\plotpoint}}
\put(463.05,344.71){\usebox{\plotpoint}}
\multiput(468,344)(20.473,-3.412){0}{\usebox{\plotpoint}}
\multiput(474,343)(19.957,-5.702){0}{\usebox{\plotpoint}}
\put(483.37,340.66){\usebox{\plotpoint}}
\multiput(488,340)(20.473,-3.412){0}{\usebox{\plotpoint}}
\multiput(494,339)(19.957,-5.702){0}{\usebox{\plotpoint}}
\put(503.68,336.55){\usebox{\plotpoint}}
\multiput(507,336)(19.957,-5.702){0}{\usebox{\plotpoint}}
\multiput(514,334)(20.473,-3.412){0}{\usebox{\plotpoint}}
\put(523.98,332.43){\usebox{\plotpoint}}
\multiput(527,332)(19.690,-6.563){0}{\usebox{\plotpoint}}
\multiput(533,330)(20.547,-2.935){0}{\usebox{\plotpoint}}
\put(544.25,328.29){\usebox{\plotpoint}}
\multiput(546,328)(19.957,-5.702){0}{\usebox{\plotpoint}}
\multiput(553,326)(20.473,-3.412){0}{\usebox{\plotpoint}}
\put(564.41,323.46){\usebox{\plotpoint}}
\multiput(566,323)(20.473,-3.412){0}{\usebox{\plotpoint}}
\multiput(572,322)(20.547,-2.935){0}{\usebox{\plotpoint}}
\put(584.64,319.12){\usebox{\plotpoint}}
\multiput(585,319)(20.547,-2.935){0}{\usebox{\plotpoint}}
\multiput(592,318)(20.473,-3.412){0}{\usebox{\plotpoint}}
\put(604.94,315.02){\usebox{\plotpoint}}
\multiput(605,315)(20.547,-2.935){0}{\usebox{\plotpoint}}
\multiput(612,314)(19.690,-6.563){0}{\usebox{\plotpoint}}
\multiput(618,312)(20.547,-2.935){0}{\usebox{\plotpoint}}
\put(625.23,310.96){\usebox{\plotpoint}}
\multiput(631,310)(19.957,-5.702){0}{\usebox{\plotpoint}}
\multiput(638,308)(20.473,-3.412){0}{\usebox{\plotpoint}}
\put(645.52,306.78){\usebox{\plotpoint}}
\multiput(651,306)(19.690,-6.563){0}{\usebox{\plotpoint}}
\multiput(657,304)(20.547,-2.935){0}{\usebox{\plotpoint}}
\put(665.73,302.42){\usebox{\plotpoint}}
\multiput(670,301)(20.547,-2.935){0}{\usebox{\plotpoint}}
\multiput(677,300)(20.473,-3.412){0}{\usebox{\plotpoint}}
\put(685.99,298.15){\usebox{\plotpoint}}
\multiput(690,297)(20.473,-3.412){0}{\usebox{\plotpoint}}
\multiput(696,296)(20.547,-2.935){0}{\usebox{\plotpoint}}
\put(706.25,293.92){\usebox{\plotpoint}}
\multiput(709,293)(20.547,-2.935){0}{\usebox{\plotpoint}}
\multiput(716,292)(19.690,-6.563){0}{\usebox{\plotpoint}}
\put(726.42,289.37){\usebox{\plotpoint}}
\multiput(729,289)(20.473,-3.412){0}{\usebox{\plotpoint}}
\multiput(735,288)(19.957,-5.702){0}{\usebox{\plotpoint}}
\put(746.74,285.32){\usebox{\plotpoint}}
\multiput(749,285)(19.690,-6.563){0}{\usebox{\plotpoint}}
\multiput(755,283)(20.547,-2.935){0}{\usebox{\plotpoint}}
\put(767.00,281.17){\usebox{\plotpoint}}
\multiput(768,281)(19.957,-5.702){0}{\usebox{\plotpoint}}
\multiput(775,279)(20.473,-3.412){0}{\usebox{\plotpoint}}
\put(787.32,277.10){\usebox{\plotpoint}}
\multiput(788,277)(19.690,-6.563){0}{\usebox{\plotpoint}}
\multiput(794,275)(20.547,-2.935){0}{\usebox{\plotpoint}}
\multiput(801,274)(19.690,-6.563){0}{\usebox{\plotpoint}}
\put(807.34,271.95){\usebox{\plotpoint}}
\multiput(814,271)(20.473,-3.412){0}{\usebox{\plotpoint}}
\multiput(820,270)(19.957,-5.702){0}{\usebox{\plotpoint}}
\put(827.66,267.89){\usebox{\plotpoint}}
\multiput(833,267)(20.547,-2.935){0}{\usebox{\plotpoint}}
\multiput(840,266)(19.690,-6.563){0}{\usebox{\plotpoint}}
\put(847.93,263.72){\usebox{\plotpoint}}
\multiput(853,263)(19.690,-6.563){0}{\usebox{\plotpoint}}
\multiput(859,261)(20.547,-2.935){0}{\usebox{\plotpoint}}
\put(866,260){\usebox{\plotpoint}}
\vspace{-10mm}
\caption{$M_{V}/e$ versus $1/N_C^{2}$ in the continuum.
The error bars are estimated  from the data for
different $r$.}
\label{fig6}
\end{figure}

\end{document}